\documentclass[aps,prl,twocolumn,showpacs,groupedaddress]{revtex4}  
\usepackage{graphicx}  
\usepackage{dcolumn}   
\usepackage{bm}        
\usepackage{amssymb}   

\begin{document}



\title{Deterministic spatio-temporal control of nano-optical fields in optical antennas and nano transmission lines}

\author{J.S.~Huang$^{1}$}\thanks{email: jhuang@physik.uni-wuerzburg.de}
\author{D.V.~Voronine$^{2}$}
\author{P.~Tuchscherer$^{2}$}
\author{T.~Brixner$^{2}$}
\author{B.~Hecht$^{1}$}

\affiliation{\vspace{0.1 in}}
\affiliation{$^{1}$Nano-Optics {\rm \&} Biophotonics Group, Physikalisches Institut, Experimentelle Physik 5 {\rm \&} R{\"o}ntgen Research Center for Complex Material Research (RCCM), Universit{\"a}t W{\"u}rzburg, Am Hubland, 97074 W{\"u}rzburg, Germany}
\affiliation{$^{2}$Institut f{\"u}r Physikalische Chemie, Universit{\"a}t W{\"u}rzburg, Am Hubland, 97074 W{\"u}rzburg, Germany}
\date{\today}

\begin{abstract}
We show that pulse shaping techniques can be applied to tailor the ultrafast temporal response of the strongly confined and enhanced optical near fields in the feed gap of resonant optical antennas (ROAs). Using finite-difference time-domain (FDTD) simulations followed by Fourier transformation, we obtain the impulse response of a nano structure in the frequency domain, which allows obtaining its temporal response to any arbitrary pulse shape. We apply the method to achieve deterministic optimal temporal field compression in ROAs with reduced symmetry and in a two-wire transmission line connected to a symmetric dipole antenna. The method described here will be of importance for experiments involving coherent control of field propagation in nanophotonic structures and of light-induced processes in nanometer scale volumes.

\end{abstract}

\maketitle

Resonant optical antennas (ROA) are metal nano structures that upon illumination with light of well-defined wavelength and polarization exhibit hot spots in which the local optical field is spatially confined and strongly enhanced due to geometry-dependent plasmon resonances \cite{Schuck2005,OAScience2005Bert}. When excited by a pulsed laser such hot spots provide the possibility to e.g.~investigate nonlinear behavior of materials at moderate pulse energies \cite{OAScience2005Bert,CuspFieldEnhancemantNL2007,HHGbyPlasmaNature2008Kim}. Vice versa, well designed ROAs are able to enhance and direct the radiation of point-like sources that drive the antenna from within a hot spot \cite{SuperEmitterPRL2005Bert,QtrLambda2007vanHulst} thus providing the basis for enhanced single photon sources \cite{SuperEmitterPRL2005Bert} and antenna-enhanced local spectroscopy \cite{MoernerSERS}. An important future application of ROAs is coherent spatio-temporal control of optical fields for antenna-enhanced local spectroscopies. Coherent control of near-fields by amplitude and polarization pulse shaping has recently been theoretically proposed \cite{M.Stockman2002,Brixner-PRL2005,Brixner-PRB2006,Seideman-JPB2007} and experimentally realized \cite{Brixner-Nature2007}. In extension of such experiments it is envisaged to perform spectroscopy of complex matter in contact with a ROA assembly, which will allow studying ultrafast temporal dynamics with nanoscale spatial resolution \cite{Brixner-unpubl2008}. The well-established techniques of femtosecond quantum control \cite{control1,control2} can be realized on the nano-scale. To reach this goal it is important to understand how the resonant character of nano antennas affects the pulse shape in the feed gap and which degrees of freedom are available to be exploited for coherent control of their near-fields. So far efficient ROAs have been realized using noble metal nano rods which are produced by microfabrication techniques. The plasmon resonances of individual rods are determined by their dimensions and dielectric function \cite{ElSayed1999} or equivalently by length-dependent Fabry-Perot resonances of a plasmon wave \cite{KrennPRL2005} propagating with an effective wavelength \cite{EffWaveLengthPRL2007Lukas} along the rod and being reflected at its ends. While a single rod may be considered a monopole antenna \cite{QtrLambda2007vanHulst}, dipole antennas are formed by aligning two nano rods end to end thus creating a very small feed gap between the two rods where optical fields are concentrated \cite{OAScience2005Bert}. The plasmon resonances of such coupled systems are found by applying the hybridization model \cite{Hybride} which for excitation with a polarization parallel to the dipole antenna predicts a splitting of the fundamental single nano rod resonance into a red-shifted and a blue-shifted hybrid mode of the coupled system. In the present Letter only the red-shifted resonance is further considered.

In this Letter we use the finite-difference time-domain method (FDTD solutions, Lumerical solutions, Inc, Cananda) to investigate the temporal response of ROA structures. To this end, we obtain the source independent impulse response spectra and corresponding spectral phases via fast Fourier transformation of the structure's temporal response to a single excitation pulse with a Gaussian envelope. The structure's temporal response to arbitrarily shaped pulses is then obtained by a multiplication of the impulse response with the excitation pulse in the frequency domain followed by inverse Fourier transformation. We show that by applying pulses with an inverse phase with respect to the impulse response optimal temporal compression is achieved. As further extension of our technique we demonstrate that laser pulse shaping can also be used to compensate and control pulse broadening due to dispersion in a nanometric plasmonic two-wire transmission line excited via a ROA \cite{JS2008}. We therefore extend the classical experiments of adaptive pulse compression \cite{Gerber-APB1997,1997Silberberg} and the concept of energy localization in random nano-structures \cite{M.Stockman2002} to the deterministic spatio-temporal control of nano-optical fields in elementary nano-photonic building blocks. 

\begin{figure}[htb]
\begin{center}
\includegraphics*[width=1.0\linewidth]{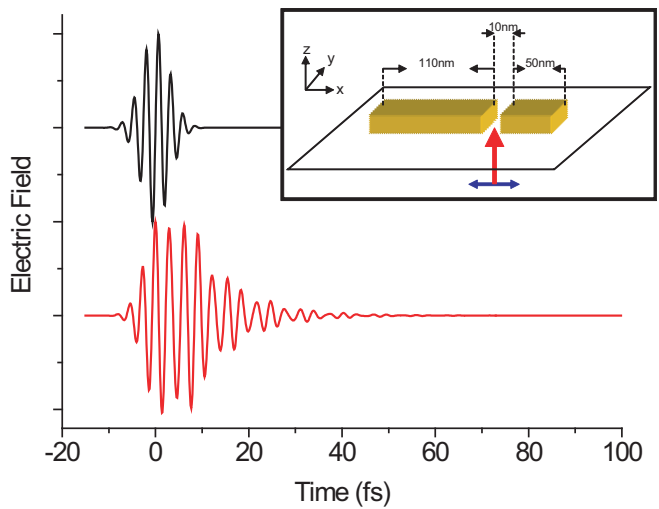}
\end{center}
\caption{Time-dependent electric fields of the default excitation source (black) and the corresponding near-field response (red) at the feed gap center of a dipole antenna with a displaced gap. Antenna dimensions are $20\times20\times170~\mbox{nm}^{3}$ including a 10 nm gap, which is shifted 30 nm away from the antenna center. The displaced gap results in arm lengths of 110 and 50 nm as sketched in the inset.}
\label{fig:1}
\end{figure}

To excite the nano structures, an x-polarized 10 fs pulse centered at 711 nm is used as a default source in the FDTD simulations (see Fig.~\ref{fig:1}). Its electric field may be expressed as
\begin{equation}\label{source}
{\bf S}(t)={\bf S}_{\rm 0} e^{\frac{1}{2}\left[\frac{t-t_{\rm 0}}{\Delta t}\right]^{2}}\sin[\omega(t-t_{0})]\; ,
\end{equation}
where ${\bf S}_{\rm 0}$ is the peak amplitude of the source electric field, $t_{\rm 0}$ is the time offset of the pulse, $\Delta t$ is the pulse duration and $\omega $ is the carrier angular frequency. The default source has a flat spectral phase (no chirp). The nano structures to be investigated are made of gold on silica substrates in vacuum. We use dielectric functions for gold and silica that are based on experimental data \cite{AuEpsilonPRB1972,SiO2EpsilonHandbook1985}. The source is injected from within the silica half space with incidence perpendicular to the gold-silica interface. Structures are discretized using mesh steps in $x$, $y$ and $z$ direction of 1~nm to obtain reliable results with reasonable simulation time. All the structures have faces and edges parallel to the rectangular mesh grid. We therefore exclude the influence of spurious resonances due to staircasing effects.


\begin{figure}[htb]
\begin{center}
\includegraphics*[width=1.07\linewidth]{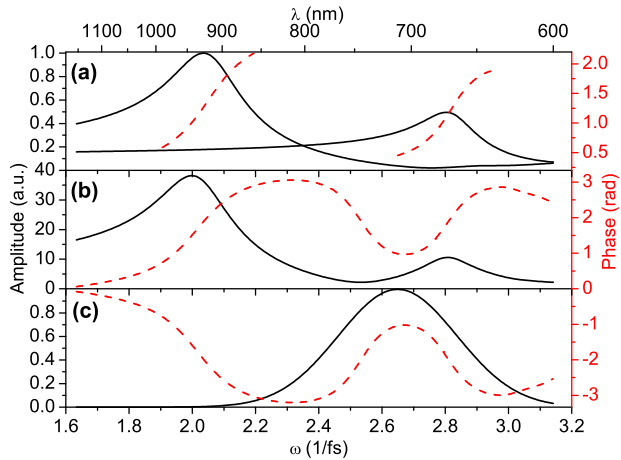}
\end{center}
\caption{Source-independent impulse spectra (black, solid) and spectral phases (red, dashed) for linearly polarized excitation along the long axis of (a) both, an isolated 50 nm and 110 nm gold nano-rod recorded at a point on the long axis of the rod, 5~nm from its end in air, (b) the respective diplaced-gap antenna consisting of the two rods shown in (a), recorded at the center of the 10~nm feed gap. (c) New phase-shaped source pulse (black, solid) with a modified spectral phase (red, dashed) which is the inverse of the diplaced-gap antenna spectral phase shown in (b).}
\label{fig:2}
\end{figure}

We first study the temporal response of a dipole antenna with a displaced gap, i.e.~an antenna consisting of two gold nano rods with the same cross section ($20\times20~\mbox{nm}^{2}$), but different length (110~nm and 50~nm, respectively) separated by a feed gap of 10~nm as depicted in the inset of Fig.~\ref{fig:1}. FDTD simulations provide the time-dependent electric fields ${\bf E}(t)$ in the vicinity of the structure, e.g.~in the feed gap (red trace in Fig.~\ref{fig:1}). Both, the source field ${\bf S}(t)$ (black trace in Fig.~\ref{fig:1}) and the corresponding response of the structure are Fourier transformed to yield the respective quantities ${\bf S}(\omega)$ and ${\bf E}(\omega)$ in the frequency domain. The so-called impulse response function for any desired point $\bf r$ in the vicinity of the structure is now obtained by normalizing the response of the structure with the source spectrum
\begin{equation}\label{normalization}
{\bf E}_{\rm impulse}(\omega,\bf r) = \frac{{\bf E}(\omega,\bf r)}{{\bf S}(\omega)} \; .
\end{equation}
Since the frequency dependence due to the finite pulse length of the source has been removed, the impulse spectrum reveals the spectral response of the structure to a $\delta$-function pulse. It therefore serves as a Green's function in the frequency domain. A higher-order nonlinear spectral phase of the impulse response directly reveals a possible chirp introduced by the nano-structure. 
In the case considered here, each individual antenna arm exhibits its own longitudinal plasmon resonance with the expected harmonic oscillator behavior of the phase. The impulse spectrum of a displaced-gap antenna shows two clear resonance peaks due to asymmetry \cite{Bratschitsch}. These two peaks around the angular frequencies 2.0 $\mbox{fs}^{-1}$(946 nm) and 2.8 $\mbox{fs}^{-1}$(671 nm) correspond to 110 nm and 50 nm single gold nanorod resonances, whose impulse spectra and phases are shown in Fig.~\ref{fig:2} (a). The longer rod resonance is shifted to the red and shows a stronger near-field enhancement due to the fact that the longitudinal plasmonic resonance is red-shifted and becomes stronger as the aspect ratio of the rod increases \cite{ElSayed1999}. We observe that for a {\em single} rod excited by linearly polarized light along the long axis (see Fig.~\ref{fig:2} (a)) and also for a {\em symmetric} dipole antenna (data not shown) the spectral phase around the resonance is linear to a good approximation \cite{NanosphereResonatorPRL1999Lamprecht}, thus, negligible chirp is introduced. This is supported by recent experiments on symmetric bow-tie antenna arrays \cite{HHGbyPlasmaNature2008Kim}.

While in view of applications this is an important finding in itself, we also observe that already the presence of a second resonance due to an asymmetric feed gap, not to speak of more complicated structures with more resonances, introduces higher-order spectral phase behavior as illustrated in Fig.~\ref{fig:2} (b). But since the impulse response of the structure is known, it is possible to deterministically pre-compensate any pulse broadening by adapting the source pulse shape such that its spectral phase is the inverse of the impulse response spectral phase as illustrated in Fig.~\ref{fig:2} (c). As a result, the response of the structure to the so-modified source exhibits a flat spectral phase, and is optimally compressed in time \cite{Gerber-APB1997,1997Silberberg}. Most importantly, the structure response to any chosen source pulse shape can be simply obtained by multiplying the impulse response with the new source in the spectral domain without re-running expensive FDTD simulations.

\begin{figure}[htb]
\begin{center}
\includegraphics*[width=1.05\linewidth]{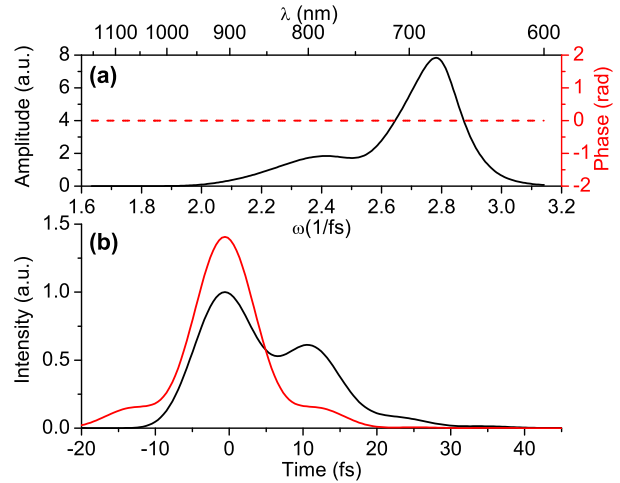}
\end{center}
\caption{ (a) New response spectrum (black solid) and spectral phase (red dashed). (b)Temporal intensity profile of the original response (black) and of the new response (red). Both intensities are normalized to the maximum intensity of the original response.}
\label{fig:3}
\end{figure}

Applying the shaped pulse to the structure, the dispersion is pre-compensated and a new response with a flat phase is obtained as shown in Fig.~\ref{fig:3} (a).
The new source excites both resonances due to the broad bandwidth of the ultrashort pulse. This results in the main peak at $\omega=2.8$ $\mbox{fs}^{-1}$ and a shoulder around $\omega=2.4$ $\mbox{fs}^{-1}$. Since the spectral phase is flat, the temporal broadening is removed. As a result, the new response is compressed in time and its duration is finally limited by the Q factor of the relevant plasmon resonance provided the excitation pulse is sufficiently short. Fig.~\ref{fig:3} (b) shows the temporal intensity profile of the well-compressed new response obtained by the inverse Fourier transformation along with the original response. The peak intensity is enhanced by a factor of 1.4. Because both resonances of the structure are excited, the compressed temporal profile in Fig.~\ref{fig:3} (b) is not a Gaussian but has shoulders. Peak intensity enhancement as observed in the present example of an asymmetric dipole antenna will be of importance for nonlinear phenomena observed in the antenna feed gap \cite{OAScience2005Bert,CuspFieldEnhancemantNL2007} or higher-order harmonic signals sensitive to the local field intensity \cite{NanosphereResonatorPRL1999Lamprecht,HHGbyPlasmaNature2008Kim}.

\begin{figure}[htb]
\begin{center}
\includegraphics*[width=1.0\linewidth]{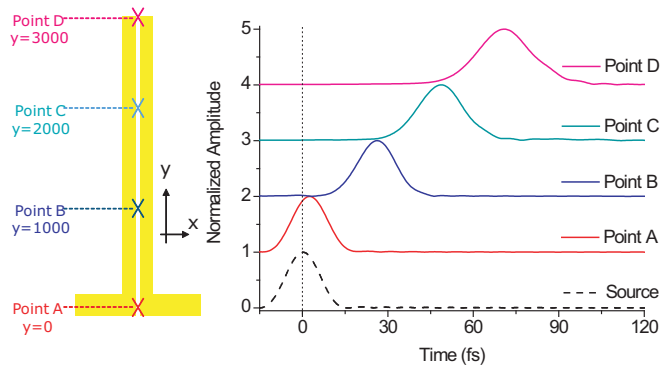}
\end{center}
\caption{Temporal profiles recorded at points A (red), B (blue), C (green) and D (pink) obtained with the default source (black dashed). Two infinitely long gold nano-wires with $20\times20\  \mbox{nm}^2$ cross-section and 10 nm separation are attached to two arms of a $20\times20\times190\ \mbox{nm}^3$ antenna. Observation point A is located at the feed gap center while points B, C and D are shifted in the +y direction from point A by 1000, 2000 and 3000 nm, respectively. Inset shows the sketch of the structure and the observation points.}
\label{fig:4}
\end{figure}

\begin{figure}[htb]
\begin{center}
\includegraphics*[width=1.05\linewidth]{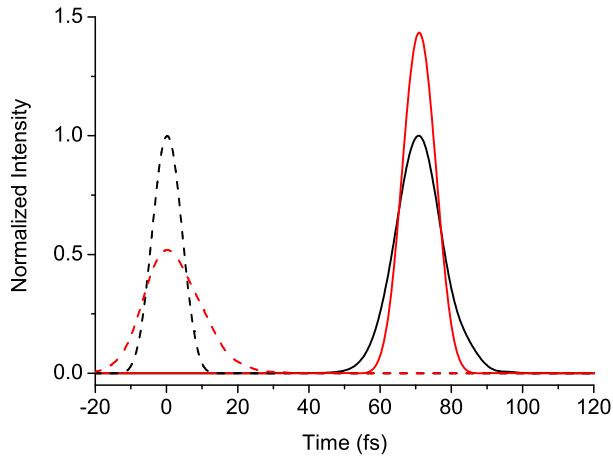}
\end{center}
\caption{Original (black) and new (red) temporal intensity profiles of the source (dashed) and the response at point D (solid). Both intensities are normalized to the maximum intensity of the original response.}
\label{fig:5}
\end{figure}

As the complexity of the investigated structures increases so does the number of resonances that determine the response of a structure to an excitation with a short laser pulse, causing a stronger temporal broadening of the near-field. As an extreme case we study a two-wire transmission line whose equivalent circuit is a chain of interacting resonant circuits each representing an infinitely short section of the transmission line \cite{FieldandWaveDKCheng}. The effect of these interacting resonant circuits is that excitations of different frequencies propagate at different velocities along the transmission line, therefore, causing dispersion. To demonstrate the ability of manipulating the temporal profile of a signal traveling down the transmission line, we attached a symmetric dipole antenna to a two-wire transmission line \cite{JS2008,Pohl2000}. The antenna is used here to efficiently excite the fundamental mode of the transmission line while introducing negligible chirp itself. The present implementation and other related structures may in future be applied to create integrated photonic circuitry for various purposes. Upon illumination of the dipole antenna, the field is first spatially confined and enhanced in the feed gap and then travels as a TE mode confined between the two wires of the transmission line. We have recently found that these nano-transmission lines at optical frequencies behave quite similiar to those in the RF regime. Concepts of classical electronic circuits, e.g.~impedance matching, are well applicable \cite{JS2008}. A propagating pulse along such transmission line is expected to be broadened by group velocity dispersion \cite{2005PRB72_PlanarSPWGFreqDispersion_Dionne} and phase modulation in analogy to effects also known from optical fibers \cite{FiberDispersionAPL1983Nikolaus}. As shown in Fig. \ref{fig:4}, the temporal responses obtained at different positions along the transmission line show broadening which is increasing with the distance traveled. This temporal broadening, along with Ohmic losses, rapidly diminishes the peak intensity of the signal. By applying a phase-shaped pulse to the incoupling antenna using the method outlined above, the signal is re-compressed at point D and the field intensity is improved by a factor of 1.5 as shown in Fig. \ref{fig:5}. Such signal compression and the resulting peak field enhancement facilitate the detection and manipulation of signals traveling in a photonic circuit. In addition to the compression at point D, the new source shows a broader temporal intensity profile and a lower peak intensity with the same pulse energy. Since the local field is highly enhanced in the feed gap and easily exceeds the damage threshold \cite{Schuck2005, OAScience2005Bert, HHGbyPlasmaNature2008Kim, Damagethreshold1998Gudde}, this broadened excitation pulse may prevent the structure from being damaged by the source. This demonstrates that pulse streching and re-compression can be applied in nanophotonics.


In conclusion, the FDTD method is useful to simulate the impulse response of plasmonic nanostructures with multiple resonances. Responses to arbitrary source pulse shapes can be obtained from a simple multiplication followed by an inverse Fourier transformation without re-running expensive FDTD simulations. The temporally broadened localized fields of an optical antenna or propagating transmission line modes can be re-compressed by deterministic shaping of the source spectral phase thus enhancing the peak intensities in the structure. This work will be of importance for future experiments involving coherent control of field propagation in nanophotonic devices and light-induced processes in nanometer scale volumes. In addition to spatio-temporal compression of local fields, the method presented here allows optimizing the temporal response of real optical antennas and more complex structures which deviate from the ideal shape and, therefore, sets the basis for future subwavelength coherent control applications.

B.H.~and J.S.H.~wish to acknowledge Thorsten Feichtner for helpful discussions and the Wolfermann-N\"ageli foundation for finacial support
T.B.~acknowledges funding by the DPG through the Emmy-Noether Program.


\end{document}